\documentstyle[12pt]{article}
\newlength{\ms}

\def\ss{\,}
\def\np#1  { Nucl. Phys. {\bf  B#1} }

\def\pl#1  { Phys. Lett.  {\bf B#1} }
\def\cmp   { Commun. Math. Phys. }

\def\ijmp#1{ Int. J. Mod. Phys. {\bf A#1} }
\def\mpl#1 { Mod. Phys. Lett. {\bf A#1} }
\def\del {\partial}
\def\chat {{\hat c}}
\def\IR{\mbox{\bf I$_R$}}
\def\S#1{\Sigma^{(#1)}}
\def\scO{{\cal O}}
\def\xtilde{{\tilde x}}
\def\Ptilde{{\tilde P}}
\def\Ctilde{{\tilde C}}
\def\Btilde{{\tilde B}}
\def\Qtop{Q_{\mbox{\scriptsize top}}}
\def\Qtoptilde{{\tilde Q}_{\mbox{\scriptsize top}}}
\def\xbos{x_{\mbox{\scriptsize b}}}
\def\ybos{y_{\mbox{\scriptsize b}}}
\def\Pbos{P_{\mbox{\scriptsize b}}}
\def\Cbos{C_{\mbox{\scriptsize b}}}
\def\Bbos{B_{\mbox{\scriptsize b}}}

\def\beq{\begin{equation}}
\def\eeq{\end{equation}}
\def\beqa{\begin{eqnarray}}
\def\eeqa{\end{eqnarray}}
\def\nn{\nonumber}
\def\disp#1{{\displaystyle #1}}

\def\parmedskip        {  \par\medskip  }
\def\parsmallskip      {  \par\smallskip  }
\def\parbigskipn        {  \par\bigskip\noindent  }

\begin{document}
\baselineskip=0.7cm
\setlength{\ms}{3mm}



\begin{titlepage}

    \begin{normalsize}
     \begin{flushright}
                 YITP-95-11\\
                 hep-th/9511186 \\
                 November 1995
     \end{flushright}
    \end{normalsize}
    \begin{LARGE}
       \vspace{1cm}
       \begin{center}
         On the equivalence of fermionic string to\\
         bosonic string in two dimensions\\
       \end{center}
    \end{LARGE}

  \vspace{5mm}

\begin{center}
           Hiroshi I{\sc shikawa}
           \footnote{E-mail address:
              hiroshi@yukawa.kyoto-u.ac.jp}
           \footnote{JSPS Research Fellow}\\
      \vspace{4mm}
        {\it Yukawa Institute for Theoretical Physics} \\
        {\it Kyoto University, Kyoto 606-01, Japan}\\
      \vspace{1cm}

    \begin{large} ABSTRACT \end{large}
        \par
\end{center}
\begin{quote}
 \begin{normalsize}
\ \ \ \
Two-dimensional fermionic string theory is shown to have a structure
of topological model, which is isomorphic to a tensor product of two
topological ghost systems independent of each other. One of them is
identified with $c=1$ bosonic string theory while the other has
trivial physical contents. This fact enables us to regard
two-dimensional fermionic string theory as an embedding of $c=1$
bosonic string theory in the moduli space of fermionic string
theories. Upon this embedding, the discrete states of $c=1$ string
theory are mapped to those of fermionic string theory,
which is considered to be the origin of the similarity between the
physical spectra of these two theories.
We also discuss a novel BRST operator associated with this
topological structure.
\end{normalsize}
\end{quote}

\end{titlepage}
\vfil\eject

One of the most significant developments in bosonic string theory in
less than two dimensions is the discovery of the topological
description.
The exact results by means of the matrix models are successfully
reproduced in terms of a certain kind of topological theory
\cite{W1}-\cite{Li}.
In particular, the integrable structure that has been
first observed in the matrix models
\cite{FKN,DVV1} is clearly understood
from the point of view of topological theory
\cite{VV,Li,DVV2}.
In a sense, the integrability of bosonic string
theory is a natural consequence of the underlying
topological theory.
This integrable structure is also derived \cite{H} within the
continuum approach by making use of the $w_{\infty}$-currents
associated with the discrete states \cite{LZ}-\cite{W2}.

The use of the continuum approach is inevitable for the case of
fermionic, or Neveu-Schwarz-Ramond, string theory, since there are no
matrix models relevant to fermionic string theory. As is well known,
the continuum approach is a hard scheme to perform direct calculation.
However, for the case of less than two dimensions, fermionic string
theory has several common features with bosonic one, such as the
existence of the discrete states \cite{IO,BMP2}.
This fact makes us expect that fermionic
string theory has a structure analogous to bosonic one
and can be treated in a way parallel to the bosonic case.
Actually, it has been shown \cite{HaI} that
fermionic sting theory in less than two dimensions has an integrable
structure similar to that observed in the bosonic case.

Having an integrable structure in fermionic string theory, it is
natural to ask what kind of topological theory is behind it.
Since the obtained structure heavily relies on the existence of the
discrete states \cite{HaI}, the relevant topological model will be
also closely related with the discrete states.
In the bosonic (or $c=1$) case, it has been pointed
out \cite{IK1} that we can regard string theory as a bosonization of
a topological model, in which the fundamental fields are realized by
the discrete states.
The extension of this result to the fermionic case will be
a first step toward the full understanding of topological
structure in fermionic string theory.
An analysis along this direction has been performed in
ref.\cite{HiI}. However, the obtained results is incomplete and
the further investigation is necessary.

In this article, we put forward the analysis in ref.\cite{HiI} and
find a complete topological structure in two-dimensional fermionic
string theory, which we call as $\chat=1$ string theory. The
structure we have found consists of two parts, both of which can be
regarded as a topological ghost system. One of them contains the
discrete operators, in particular the ground ring generator
\cite{W2,BMP2}, as the
fundamental field, and can be identified with the structure observed
in $c=1$ bosonic string theory. This fact enables us to regard
$\chat=1$ fermionic string theory as a direct product of $c=1$
bosonic string theory and a decoupled topological sector.
Since this topological sector turns out to be unphysical, we come to
an interesting result that $\chat=1$ string theory is equivalent to
$c=1$ bosonic string theory, which we consider as the origin of the
observed similarity between these two string theories.

\parmedskip
$\chat=1$ string theory has two-dimensional target space
parameterized by two string coordinates. We denote them as
$(\phi^M,\psi^M)$ and
$(\phi^L,\psi^L)$. In the context of non-critical string theory
\cite{DLK},
$(\phi^L,\psi^L)$ corresponds to the super Liouville field and
$(\phi^M,\psi^M)$ realizes a $\chat=1$ superconformal matter. These
fields satisfy the following operator product expansion
(OPE)\footnote{Throughout this letter, we consider only the
holomorphic part of the theory.}
\beq
  \phi^i (z) \phi^j (w) \sim -\delta^{ij}\ln(z-w) \ss , \ss
  \psi^i (z) \psi^i (w) \sim \frac{\delta^{ij}}{z-w} \ss, \ss\ss
  i,j = L,M \ss.
\eeq
The associated stress tensor and supercurrent are written as
\begin{eqnarray}
  T^i &=& -{1 \over 2} (\del \phi^i)^2 + i \lambda^i \del^2 \phi^i -
{1 \over 2} \psi^i \del \psi^i \, , \\
  \label{super}
  G^i &=& i \del \phi^i \psi^i + 2\lambda^i \del \psi^i \, ,
\end{eqnarray}
where $\lambda^M=0$ and $\lambda^L=-i$.
In order to define the BRST operator, we need to introduce
the fermionic $(b,c)$ and the bosonic $(\beta,\gamma)$
ghosts. The stress tensor and the supercurrent for the ghost sector
take the form
\beqa
  T^G &=& -2 \del b\ss c - \del b\ss c -
  {3 \over 2}\del \beta\ss\gamma -
  {1 \over 2} \beta\del\gamma \, , \\
  G^G &=& b\gamma - 3 \del c\ss\beta - 2 c\del\beta  \, ,
\eeqa
where we take the signature for the bosonic ghost as
$\gamma(z) \beta(w) \sim 1/(z-w)$.
Using these fields,
the BRST operator $Q_{\chat=1}$ of $\chat=1$ string theory is
written as
\beqa
\label{BRST}
  Q_{\chat=1} &=& \oint\! dz\, j_{BRST}(z) \nonumber \\
  &=& \oint\! dz \left(
        c (T^M + T^L) - \frac{1}{2} \gamma (G^M + G^L) + b c \del c
                - \frac{1}{4} b \gamma^2
        + \frac{1}{2} \del c \, \beta \gamma - c \beta \del \gamma
                \right)  .
\eeqa

As usual, we have to bosonize the fermions and the bosonic ghosts in
order to treat the different pictures together \cite{FMS}.
Our convention is as follows:
\beqa
\label{bosonization}
& &\psi^{\pm}=e^{\pm ih}\,\mbox{\bf I}^{\pm 1}, \nn\\
& &\gamma=\eta e^{u}\,\mbox{\bf I},\qquad\quad
\beta=\del\xi\ss e^{-u}\,\mbox{\bf I}^{-1}, \\
& &h(z)h(w)\;\; \sim\;\; -\ln(z-w) \;\;\sim \;\;u(z)u(w).\nn
\eeqa
In the above equation,
$\psi^{\pm}={1 \over \!\!\sqrt{2}}\left(\psi^M\pm
i\psi^L\right)$. $h$ and $u$ are free bosons, and ($\xi,\eta$)
is a pair of fermionic ghosts with spin (0,1).
{\bf I} is a cocycle factor which anti-commutes with $bc$-ghosts and
commutes with the other fields.
The spin field that creates the Ramond ground state from the
SL(2,$\bf{C}$) invariant vacuum is expressed as
$e^{-{ih\over 2}}e^{-{u \over 2}}\,\mbox{\bf I$_R$}$.
Here, the cocycle factor {\bf I$_R$} is introduced so that the spin
field has definite (bosonic) statistics with respect to the BRST
operator (\ref{BRST}). The explicit form of {\bf I$_R$} is given
by
\beq
\mbox{\bf I$_R$}=e^{{\pi i \over 2}(N_h-N_u)},
\eeq
where $N_h={1 \over 2\pi i}\oint\! dz\, i\del h$ and
$N_{u}=-{1\over 2\pi i}\oint\! dz\,\del u$.
For the vertex operators
$\S{p,q}=e^{iph + qu}$, this cocycle factor acts as follows
\beq
  \IR\ss \S{p,q}\ss \IR^{-1} = e^{\frac{\pi}{2}(p-q)i}\ss\S{p,q} \ss .
\eeq

As is shown in ref.\cite{BLNW}, any fermionic string theory
containing a $U(1)$ current has a twisted $N=3$ superconformal algebra
that involves an improved BRST current as one of the generators.
$\chat=1$ string theory also has an $N=3$ algebra, since there are
two $U(1)$ currents $\del\phi^{M,L}$ in the theory.
The $N=2$ part of this algebra is given as follows
\beq
\label{N=2}
\begin{array}{rcl}
  T &=& T^M + T^L + T^G \ss , \\[\ms]
  J &=& {\displaystyle
       cb - \beta\gamma
       -\frac{1}{\sqrt{2}}( i \del X^+ + q i \del X^-)\ss, }\\[\ms]
  G^+ &=& j_{BRST} \nn \\
  & &\mbox{ } {\displaystyle + \del\!\left(
  \frac{1}{2}c\beta\gamma +
  \frac{1}{\sqrt{2}} c (i\del X^+ + q i \del X^- ) +
  \frac{1}{\sqrt{2}}\del c -
  \frac{1}{2\sqrt{2}}\gamma (\psi^+ + q \psi^- )
  \right) \ss, }\\[\ms]
  G^- &=& b \ss .
\end{array}
\eeq
For the rest of the $N=3$ generators, see ref.\cite{HiI}.
Here, $q$ is a free parameter and
$X^{\pm} = \frac{1}{\sqrt{2}}(\phi^M \pm i \phi^L)$.
The central charge of this algebra is
$3 q$.

The main result of our analysis is the fact that the $N=2$
superconformal algebra (\ref{N=2}) in $\chat=1$ string theory can
be expressed in the following form
\beq
\label{top_structure}
\begin{array}{rccc}
  T &=& \del x\ss P + \del C\ss B  &\!\!\!\!\!\!
        {\displaystyle
        -\frac{1}{2}\xtilde\del\Ptilde
        +\frac{1}{2}\del\xtilde\ss\Ptilde
        -\frac{1}{2}\Ctilde\del\Btilde
        +\frac{1}{2}\del\Ctilde\ss\Btilde}\ss ,\\[\ms]
  J &=& {\displaystyle
    \frac{1}{2}(1 - q)x P +
    \frac{1}{2}(1 + q) C B } &\!\!\!\!\!\!
    {\displaystyle
    -\frac{1}{2}\xtilde\Ptilde
    -\frac{1}{2}\Ctilde\Btilde }\ss , \\[\ms]
  G^+ &=& {\displaystyle
    -\frac{1}{2}(1-q)\del C\ss x
    + \frac{1}{2}(1 + q) C \del x } & \!\!\!\!\!\!
    \mbox{ } + \Btilde\Ptilde \ss , \\[\ms]
  G^- &=& B P & \!\!\!\!\!\!
    {\displaystyle
    -\frac{1}{2}\del\Ctilde\ss\xtilde
    +\frac{1}{2}\Ctilde\del\xtilde } \ss .
\end{array}
\eeq
Here, we introduced two sets of fields
$(x,P,C,B)$ and $(\xtilde,\Ptilde,\Ctilde,\Btilde)$.
One of them is defined as
\beq
\label{set1}
\begin{array}{rcl}
x &=& \disp{\left[(cb + \sqrt{2} i \del X^-)\S{1/2,1/2} +
      \frac{1}{\sqrt{2}}b\eta\, \S{-1/2,3/2} \right.} \\
  && \quad\quad\quad\quad  \disp{\left. \frac{ }{ }
      -\sqrt{2} c\del\xi\,\S{-1/2,-1/2} +
      \S{-3/2,1/2} \right]
      e^{\frac{1}{\sqrt{2}}i X^+} \IR^{-1} }\ss, \\[\ms]
P &=& \disp{\S{-1/2,-1/2} e^{-\frac{1}{\sqrt{2}}i X^+}\IR}
       \ss , \\[\ms]
C &=&  \disp{c\, \S{-1/2,-1/2} e^{-\frac{1}{\sqrt{2}}i X^+} \IR}
       \ss , \\[\ms]
B &=& \disp{\left(b\,\S{1/2,1/2} - \sqrt{2}\del\xi\,\S{-1/2,-1/2}
     \right) e^{\frac{1}{\sqrt{2}}i X^+} \IR^{-1}  }\ss .
\end{array}
\eeq
$x$ and $C$ have dimension $0$ whereas $P$ and $B$ has $1$.
It should be noted that $x$ and $C$ belong to the physical spectrum of
$\chat=1$ string theory \cite{IO,BMP2}. In particular, $x$ is one of
the ground ring generator in $1/2$-picture\footnote{We define the
picture charge as
$\oint\! dz (\xi\eta -\del u )$.} \cite{BMP2,HiI}.
The other set is defined as
\beq
\label{set2}
\begin{array}{rcl}
\xtilde &=& \disp{-2\sqrt{2} \S{-1/2,-1/2} \IR \ss }, \\[\ms]
\Ptilde &=& \disp{-\frac{1}{2\sqrt{2}}
      \left[-(\xi\eta - \sqrt{2} i \del X^-
           + i\del h - \del u )\S{1/2,1/2} +
      \frac{1}{\sqrt{2}}b\eta \S{-1/2,3/2} \right.} \\
  &&   \disp{\left. \frac{ }{ }
      -2\sqrt{2}
      \left(c\del\xi + \frac{1}{2}\del c\ss\xi
       -\frac{1}{2} c\xi(i\del h + \del u)\right)\S{-1/2,-1/2} +
      \S{-3/2,1/2} \right]  \IR^{-1} \ss } , \\[\ms]
\Ctilde &=& \disp{\xi \S{-1/2,-1/2} \IR^{-1} \ss ,} \\[\ms]
\Btilde &=& \disp{\left(\eta \S{1/2,1/2} +
        \sqrt{2}
        (c(i\del h + \del u) - \del c) \S{-1/2,-1/2} \right)
        \IR } \ss .
\end{array}
\eeq
These fields have dimension $1/2$.
As is seen from the presence of the cocycle factor $\IR$, all the
fields in eqs.(\ref{set1}) and (\ref{set2}) are in the Ramond sector.

We can consider each of these two sets as a realization of  a
topological ghost system, {\it i.e.}, a supersymmetric ghost
system with a twisted
$N=2$ structure.
Actually, we can check that the above fields satisfy the following
OPE's
\beq
\label{OPE}
  x(z) P(w) \sim -\frac{1}{z-w} \sim -P(z) x(w)
  \ss ,\quad
  C(z) B(w) \sim \frac{1}{z-w} \sim B(z) C(w) \ss ,
\eeq
which mean that $(x,P)$ realize a bosonic ghost while $(C,B)$ does
a fermionic one.
The tilded fields $(\xtilde,\Ptilde,\Ctilde,\Btilde)$ satisfy the
same OPE's as above and commute with the fields in eq.(\ref{set1}).
Hence, we can consider that these two sets of fields,
(\ref{set1}) and (\ref{set2}), realize two supersymmetric ghost
systems independent of each other.
The $N=2$ structure of these two is determined as displayed in
eq.(\ref{top_structure}).
The central charge takes $3q$ and $0$, respectively.

Putting all these facts together, we can conclude that $\chat=1$
string theory can be viewed as a direct product of two topological
ghost systems, $(x,P,C,B)$ and
$(\xtilde,\Ptilde,\Ctilde,\Btilde)$.
In fact, the discrete states of $\chat=1$ string theory can be
obtained as the physical states of this topological system. The
physical spectrum of the ghost system is determined by the BRST
operators,
$\Qtop$ and
$\Qtoptilde$, defined in eq.(\ref{top_structure}):
\beq
  \Qtop = \oint\! dz\, C\del x \quad , \quad
  \Qtoptilde = \oint\! dz\, \Btilde \Ptilde \ss.
\eeq
The BRST operator of $\chat=1$ string theory is expressed as
$Q_{\chat=1} = \Qtop + \Qtoptilde$.
Since both of $\Qtop$ and $\Qtoptilde$ are written in bilinear form of
the fundamental fields, almost all the states of the ghost
system become unphysical. In particular, for the tilded system, all
the fields,
$\xtilde, \Ptilde, \Ctilde, \Btilde$, form BRST-doublets
to leave the trivial physical spectrum. On the other hand, $\Qtop$
involves one derivative and the zero mode of $C$ and $x$ is missing in
the BRST operator.
Consequently, for the non-tilded system, the zero mode of $C$ and $x$
becomes BRST-singlet and yields physical states. The physical spectrum
is generated by these two zero modes and takes the form
\beq
\label{spectrum}
  x^n \ss\ss, \ss\ss C x^n \ss , \ss\ss n = 0,1,2,\cdots \ss .
\eeq
The entire physical spectrum is a direct product of those for the two
ghost systems. Since the tilded system has trivial physical contents,
the spectrum for the total system coincides with that for the
non-tilded system given in eq.(\ref{spectrum}).
As is noticed before, both of $x$ and $C$ are realized by the discrete
operators in $\chat=1$ string theory. Using this fact and the ring
structure of physical operators, we can confirm that the physical
spectrum (\ref{spectrum}) of the ghost system reproduces the discrete
states of $\chat=1$ string theory. Of course, eq.(\ref{spectrum}) do
not cover all the discrete states. It is highly possible that the
rest is obtained by the picture-changing operation \cite{FMS}
associated with the expression in eqs.(\ref{set1}) and (\ref{set2}).
We will come to this point later.

We have seen that the physical contents of $\chat=1$ string theory is
governed by the topological system consisting of $(x, P, C, B)$ with
the BRST operator $\oint C\del x$.
Amazingly, this structure is exactly the same as that observed in
$c=1$ bosonic string theory \cite{IK1}.
To be precise, the twisted $N=2$ structure in $c=1$ string theory,
which contains the improved BRST current as $G^+$, can be expressed in
the same way as that for the
$\chat=1$ case (\ref{top_structure}). Namely, introducing an
appropriate set of fields $(\xbos,\Pbos,\Cbos,\Bbos)$\footnote{The
subscript `b' stands for `bosonic'.} that realizes a topological ghost
system, we can rewrite the $N=2$ algebra in $c=1$ string theory as
follows
\cite{IK1}:
\beq
\begin{array}{rcl}
  T &=& \del \xbos\ss \Pbos + \del \Cbos\ss \Bbos \ss ,\\[\ms]
  J &=& {\displaystyle
    \frac{1}{2}(1 - q)\xbos \Pbos +
    \frac{1}{2}(1 + q) \Cbos \Bbos } \ss , \\[\ms]
  G^+ &=& {\displaystyle
    -\frac{1}{2}(1-q)\del \Cbos\ss \xbos
    + \frac{1}{2}(1 + q) \Cbos \del \xbos }  \ss , \\[\ms]
  G^- &=& \Bbos \Pbos  \ss .
\end{array}
\eeq
Again, $q$ is a free parameter,
and both of $\xbos$ and $\Cbos$ are realized by the physical operators
of $c=1$ string theory.
In particular, $\xbos$ is one of the ground
ring generator.
The explicit form of these fields is given by
\beq
\label{bos_field}
\begin{array}{rclcrcl}
  \xbos &=& \disp{( cb + i\, \del X^- ) e^{i X^+}} \, ,&\quad\quad&
  \Cbos &=& \disp{c \, e^{-i X^+ }\, ,}\nn\\[\ms]
  \Pbos &=& \disp{e^{- i X^+ }\, ,} &\quad\quad&
  \Bbos &=& \disp{b \, e^{i X^+ } \, ,} \nn
\end{array}
\eeq
where $(c,b)$ is the diffeomorphism ghost in $c=1$ string theory and
$X^\pm = \frac{1}{\sqrt{2}}(X \pm i\phi)$ is the light-cone
combination of string coordinates
(Note that this is exactly the same form as the
supersymmetric bosonization of ghost systems \cite{super_bos}). The
discrete states can be regarded as the physical states of this
topological system in the same manner as the
$\chat=1$ case described above.
One of the ground ring generator other than $\xbos$, which we denote
as $\ybos = (cb - i\del X^+) e^{-i X^-}$, works as the
picture-changing operator of this system \cite{IK1}.
In this way, $c=1$ string theory can
also be viewed as a realization of a topological ghost system, which is
equivalent to
$(x,P,C,B)$ appearing in $\chat=1$ string theory.

Through this structure, we can identify $\chat=1$ fermionic string
theory with $c=1$ bosonic theory.
Namely, we can perform the following successive identifications:
\beq
\begin{array}{rcccl}
  \chat=1 &\sim &
      (x,P,C,B) \otimes (\xtilde,\Ptilde,\Ctilde,\Btilde) & & \\[\ms]
       &\sim & (x,P,C,B) & \\[\ms]
       &\sim & (\xbos,\Pbos,\Cbos,\Bbos) &
       \sim & c=1 \ss .
\end{array}
\eeq
In the last line, we identified the fields (\ref{bos_field}) in $c=1$
string theory with its counterpart in $\chat=1$ string theory.  In a
sense,
$c=1$ bosonic string theory is embedded in the space of fermionic
string theories.
Schematically,
\beq
\label{embedding}
  c=1 \quad\rightarrow\quad c=1 \otimes \mbox{topological sector}
  \quad\sim\quad \chat =1 \ss .
\eeq
The apparent similarity between $c=1$ and $\chat=1$ theories can be
considered as the result of this structure.

The embedding in eq.(\ref{embedding}) reminds us another embedding
found by Berkovits and Vafa \cite{BV}.
They have shown that any vacuum of bosonic string
theory can be regarded as a special type of vacuum of fermionic string
theory.
The essential point of their argument is that it is possible to
construct an
$N=1$ superconformal algebra with $c=15$ by combining a $c=26$ Virasoro
algebra with a pair of fermionic ghosts.
The resultant fermionic string vacuum, which we call as the
Berkovits-Vafa (BV) vacuum, is argued to be equivalent to the original
bosonic one \cite{BV}-\cite{IK2}.
Especially, it has been shown \cite{IK2} that the BV vacuum can be
regarded as a direct product of the original bosonic vacuum with a
trivial topological sector
\beq
\label{BV}
  \mbox{BV vacuum} \sim \mbox{bosonic vacuum} \otimes
  \mbox{topological sector} \ss ,
\eeq
which means that the BV vacuum also provides us with an embedding of
bosonic string vacua.

Thus, we have two types of embedding for $c=1$ string theory. One is
that given in eq.(\ref{embedding}) and produces $\chat=1$ string
theory. The other is the BV vacuum (\ref{BV}), in which the bosonic
vacuum is taken as $c=1$ string theory.
There are some differences between these two, and we should consider
them as inequivalent embeddings.
Indeed, the $N=1$ superconformal structure of fermionic string theory
is totally different for these two.
In $\chat=1$ string theory, the $N=1$ superconformal symmetry on the
worldsheet is realized linearly (see eq.(\ref{super})).
On the other hand, the BV vacuum has a
non-linear supersymmetry, which is associated with a breaking of
superconformal symmetry to conformal one \cite{K,M}.
Since the physical implication of this non-linear symmetry is not
clear, it is worthwhile to compare these two embeddings in order to
make definite the status of non-linear realized
symmetry in string theory.

Before concluding this article, we consider the issue of
picture-changing operation in the ghost systems defined in
eqs.(\ref{set1}) and (\ref{set2}).
As is well known, bosonization of bosonic ghost system enlarges the
Fock space and enables us to treat vacua with different sea-level,
{\it i.e.}, different picture, together.
So, we need a careful analysis in order to obtain the correct physical
spectrum of the model in the entire Fock space.
This task is relatively easy
for the ordinary bosonization such as adopted in ref.\cite{FMS}.
In contrast to that, the realization of the ghost
system in the present case is much complicated (see eqs.(\ref{set1}),
(\ref{set2})) and awkward to perform calculations.
However, this complication is just an artifact
and can be removed by the following similarity transformation
$\scO \rightarrow e^R \scO e^{-R}$, where $R$ is defined as
\beq
  R = -\sqrt{2} c \del \xi\ss e^{-ih - u}
  = -\sqrt{2} c\psi^- \beta \ss .
\eeq
After this transformation,
the fundamental fields, (\ref{set1}), (\ref{set2}), of the ghost
system turn into the following form
\beqa
\label{set1mod}
&&\!\!\!\!
\begin{array}{rcl}
x &=& \disp{\left[(cb + \sqrt{2} i \del X^-)\S{1/2,1/2} +
      \frac{1}{\sqrt{2}}b\eta \S{-1/2,3/2} \right]
      e^{\frac{1}{\sqrt{2}}i X^+} \IR^{-1} }\ss, \\[\ms]
P &=& \disp{\S{-1/2,-1/2} e^{-\frac{1}{\sqrt{2}}i X^+}\IR}
       \ss , \\[\ms]
C &=&  \disp{c \S{-1/2,-1/2} e^{-\frac{1}{\sqrt{2}}i X^+} \IR}
       \ss , \\[\ms]
B &=& \disp{b \S{1/2,1/2}
        e^{\frac{1}{\sqrt{2}}i X^+} \IR^{-1}  }\ss ,
\end{array} \\
&& \mbox{ } \nn \\[\ms]
\label{set2mod}
&&\!\!\!\!
\begin{array}{rcl}
\xtilde &=& \disp{-2\sqrt{2} \S{-1/2,-1/2} \IR \ss }, \\[\ms]
\Ptilde &=& \disp{-\frac{1}{2\sqrt{2}}
      \left[-(\xi\eta - \sqrt{2} i \del X^-
           + i\del h - \del u )\S{1/2,1/2} +
      \frac{1}{\sqrt{2}}b\eta \S{-1/2,3/2}
      \right]  \IR^{-1}  } ,\\[\ms]
\Ctilde &=& \disp{\xi \S{-1/2,-1/2} \IR^{-1},} \\[\ms]
\Btilde &=& \disp{\eta \S{1/2,1/2} \IR } \ss .
\end{array}
\eeqa
Clearly, the above form of fields is much simpler than the original
one.
Moreover, this is almost the same form as that for the fields
(\ref{bos_field}) in $c=1$ string theory. Necessary identification is
\beq
\label{map}
\begin{array}{ccc}
  c=1   &                & \chat=1  \\[\ms]
  (c,b) &\longrightarrow & (c,b)  \\[\ms]
  X^+   &\longrightarrow &
        \disp{\frac{1}{2}(h-iu) + \frac{1}{\sqrt{2}} X^+} \\[\ms]
  X^-   &\longrightarrow & \sqrt{2} X^- \ss.
\end{array}
\eeq
Upon this identification, the set of fields,
$(\xbos,\Pbos,\Cbos,\Bbos)$, is mapped to $(x,P,C,B)$ in
eq.(\ref{set1mod}), if we neglect the second term of $x$.

This fact suggests that the genuine fundamental fields of
the topological structure in $\chat=1$ string theory are not
the fields in eqs.(\ref{set1mod}) and (\ref{set2mod}) but those
without the second term of $x$ and $\Ptilde$.
Let us denote them by attaching the subscript `0' to the original
ones:
\beqa
\label{set1mod2}
&&
\begin{array}{rcl}
x_0 &=& \disp{(cb + \sqrt{2} i \del X^-)\S{1/2,1/2}
      e^{\frac{1}{\sqrt{2}}i X^+} \IR^{-1} }\ss, \\[\ms]
P_0 &=& \disp{\S{-1/2,-1/2} e^{-\frac{1}{\sqrt{2}}i X^+}\IR}
       \ss , \\[\ms]
C_0 &=&  \disp{c \S{-1/2,-1/2} e^{-\frac{1}{\sqrt{2}}i X^+} \IR}
       \ss , \\[\ms]
B_0 &=& \disp{b \S{1/2,1/2}
        e^{\frac{1}{\sqrt{2}}i X^+} \IR^{-1}  }\ss ,
\end{array} \\
&& \mbox{ } \nn \\[\ms]
&&
\label{set2mod2}
\begin{array}{rcl}
\xtilde_0 &=& \disp{-2\sqrt{2} \S{-1/2,-1/2} \IR \ss }, \\[\ms]
\Ptilde_0 &=& \disp{\frac{1}{2\sqrt{2}}
      (\xi\eta - \sqrt{2} i \del X^-
           + i\del h - \del u )\S{1/2,1/2}  \IR^{-1} \ss }
      ,\\[\ms]
\Ctilde_0 &=& \disp{\xi \S{-1/2,-1/2} \IR^{-1} \ss ,} \\[\ms]
\Btilde_0 &=& \disp{\eta \S{1/2,1/2} \IR } \ss .
\end{array}
\eeqa
We can consider that these modified fields still realize
two topological ghost systems, since they satisfy the same OPE's as
displayed in eq.(\ref{OPE}).
Furthermore, this modification does not affect the $N=2$ structure
(\ref{top_structure})
realized in the $\chat=1$ Fock space, except for $G^+$. Namely,
substitution of the above fields for the fields in
eq.(\ref{top_structure}) do not alter the form of $T, J$ and $G^-$.
However,
the BRST current $G^+$ suffers a modification, which is written in the
form of charge as
\beqa
\label{newBRST}
  Q_0 \equiv \oint\! dz (C_0 \del x_0 + \Btilde_0 \Ptilde_0 )
  &=&  \oint\! dz (C \del x + \Btilde \Ptilde ) + Q' \\
  &=& Q_{\chat=1} + Q' \ss .\nn
\eeqa
Here $Q'$ is defined as
\beqa
  Q' &=& \oint\! dz \left[
  \frac{1}{2}\gamma \left(\psi^- i\del X^+ +
     \sqrt{2}\del \psi^-\right)
     + \frac{1}{4} b \gamma^2 \right] \ss, \\
     &=& e^{-R}\left[\oint\! dz \left(
  \frac{1}{2}\gamma \psi^- (i\del X^+ - \sqrt{2} cb)
     + \frac{1}{4} b \gamma^2
     + \frac{1}{2\sqrt{2}}\psi^-\beta\gamma^2
     \right) \right] e^R  \ss, \nn
\eeqa
and anti-commutes with $Q_{\chat=1}$ and $Q'$ itself
\beq
  \{Q',Q_{\chat=1}\} = \{Q',Q'\} = 0 \ss .
\eeq
This feature enables us to regard $Q'$ as a BRST operator independent
of $Q_{\chat=1}$. From this point of view, eq.(\ref{newBRST}) can be
seen as a statement that a topological structure, which is naturally
identified with a supersymmetric bosonization of topological ghost
system, emerges from $\chat=1$ string theory by imposing a further
restriction, or BRST, on it\footnote{
The notion that we need an additional BRST operator in order to
obtain a topological model is also discussed in ref.\cite{DK}.}.

The analysis of the physical spectrum is easily performed for $Q_0$.
The spectrum in the standard picture is spanned by the operators $x_0$
and
$C_0$ in the same way as before. In addition to this, we have two
picture-changing operators corresponding to two bosonized ghost
systems. One is that for the $(x_0,P_0,C_0,B_0)$-system and takes the
form
\beq
  y = e^{-R} \left(
     cb - \frac{1}{\sqrt{2}}i\del X^+
     - \frac{1}{2}(i\del h + \del u) \right) e^R \ss .
\eeq
This is obtained by applying the map (\ref{map}) to the ground ring
generator $\ybos$ in $c=1$ string theory.
The other picture-changing operator is that for the
$(\xtilde_0,\Ptilde_0,\Ctilde_0.\Btilde_0)$. Since the field $\xi$ is
contained only in the
$(\xtilde_0,\Ptilde_0,\Ctilde_0.\Btilde_0)$-system (see
eq.(\ref{set1mod2})), the picture-changing operator can be obtained
by the standard procedure \cite{FMS} and written as
$\{Q_0,\xi\}$.
Of course, there is no reason to expect that the physical spectrum
spanned by these operators coincides with that for $\chat=1$ string
theory. However, we can show that the physical operators with respect
to $Q_0$ is equivalent to those of $\chat=1$ string theory up to
$Q'$-exact terms:
\beqa
  x_0 &=& x + \{Q',*\} \ss ,\nn \\
  C_0 &=& C \ss , \\
  \{Q_0,\xi\} &=& \{Q_{\chat=1},\xi\} + \{Q',\xi\} \nn \ss.
\eeqa
By direct calculation, we can also show that the operator $y$ is
nothing but the ground ring generator other than $x$ in 0-picture.
This fact together with the simple form of the fields in
eqs.(\ref{set1mod2}) and (\ref{set2mod2}) seems to support the
necessity of the additional BRST operator $Q'$.
This point deserves further investigation.

\parmedskip
In this article, we have found that $\chat=1$ fermionic string theory
has a structure of topological model, which can be viewed as a direct
product of $c=1$ bosonic string theory and a trivial topological
sector. In ref.\cite{HiI}, the latter part is not recognized.
Instead, it is shown that the $(x,P,C,B)$-system (\ref{set1}) is
sufficient to reproduce the $N=2$ structure of $\chat=1$ string
theory, if we modify the derivative, or equivalently, the
stress tensor.
{}From the current point of view, the inclusion of the
$(\xtilde,\Ptilde,\Ctilde,\Btilde)$-system plays the same role as the
modification of the derivative. Since this modification is
shown \cite{HiI} to be closely related to the $N=3$ superconformal
structure in $\chat=1$ string theory, there may be some connection
between the $N=3$ algebra and our construction of the topological
structure.

It is natural to expect that our result is extended to the case of
$N=2$ fermionic string theory. If this is the case, {\it i.e.}, $N=2$
string is equivalent to $N=1$ string in two dimensions, it may
suggest that two-dimensional string theory is, in a sense, unique,
regardless of supersymmetry on the worldsheet.
The extension to the case of $W$-string is also interesting.

\parbigskipn\parsmallskip
I would like to thank S. Hirano and M. Kato for helpful discussions.
I would also like to thank Y. Matsuo for discussion and reading of
the manuscript.
This work is supported by
the Japan Society for the Promotion of Science, and
the Ministry of Education, Science and Culture.

\end{document}